\documentclass[12pt]{article}
\begin{document}
\begin{titlepage}
\vskip 0.3cm

\centerline{\large \bf Glueball Masses in Relativistic Potential Model}

\vskip 0.7cm

\centerline{A. Shpenik$^{\ast}$, Yu. Fekete$^{\dagger}$, J.
Kis$^{\ddagger}$}

\vskip .3cm

\centerline{\sl Uzhgorod State University,}
\centerline{\sl
Department of Theoretical Physics, Voloshin str. 32,}
\centerline{\sl 88000 Uzhgorod, Ukraine}

\vskip 1.5cm

\begin{abstract}
The problem of glueball mass spectra using the relativistic Dirac
equation is studied. Also the Breit-Fermi approach used to
obtaining hyperfine splitting in glueballs. Our approach is based
on the assumption, that the nature and the forces between two
gluons are the short-range. We were to calculate the glueball
masses with used screened potential.
\end{abstract}

\vskip 12cm

\hrule

\vskip .9cm

\noindent
\vfill
$ \begin{array}{ll} ^{\ast}\mbox{{\it e-mail
address:}} &
 \mbox{shpenik@org.iep.uzhgorod.ua}
\end{array}
$

$ \begin{array}{ll} ^{\dagger}\mbox{{\it e-mail address:}} &
 \mbox{kishs@univ.uzhgorod.ua}
\end{array}
$

\vfill
\end{titlepage}
\eject \baselineskip=14pt

\section{Introduction}

The existence of gluon self-coupling in QCD suggests that, in
addition to the conventional qq states, there may be non-qq mesons
- bound states including gluons (glueballs and qqg hybrids). The
existence of glueball states made from gluons is one of the
important predictions of QCD. The discovery of these glueball
states would be the strong support to the QCD theory. Therefore
the search and identification of glueballs have been a very
attractive research task. The abundance of qq mesons and possible
mixing of glueballs and ordinary mesons make the current situation
with the identification of glueball states rather complicated. So
the spectrum of QCD is expected to contain glueballs hybrids and
multiquark states. However, the theoretical quittance on the
properties of unusual states is often contradictory, and models
that agree in qq sector differ in their predictions about new
states. Moreover, the abundance of qq meson states in the region
$1-2GeV$ and glueball-quarkonium mixing makes the identification
of the lightest non-qq mesons difficult. Whereas there is a
general agreement that the lightest glueball should have quantum
numbers $0^{++}$ its mass remains controversial. The one of the
possibilities to distinguish between possibilities is to study of
production rates and decays branching ratios which are expected to
be sensitive to the constituent structure. Glueballs are
preferentially expected in production of gluon rich environment
such as $J/\Psi $ - , central production in hadron - hadron
collisions by double Pomeron exchange and pp annihilation into
mesons near threshold, or jj collision.

The quark model predicts that there are two mesons with $I=0$ in
the $^{3}P_{0}$ qq nonet but apart from the states $f_{j}\left(
1710\right) $ with $J=0$ or (and) $2$, four states $f_{0}\left(
400-1200\right) $, $f_{0}\left( 980\right) $, $f\left( 1370\right)
$ and $f_{0}\left( 1500\right) $ are listed by Particle Data
Group. There are too many controversies about these states. One of
the possibilities is to put aside the $f_{0}\left( 400-1200\right)
$ and $f_{0}\left( 980\right) $ and focus on $f\left( 1370\right)
$, $f_{0}\left( 1500\right) $, $f_{j}\left( 1710\right) $ although
the spin parity of the $f_{j}\left( 1710\right) $ is
$J^{PC}=0^{++}$ or (and) $2^{++}$ controversial. Recently several
authors study the quarkonia-glueball contents of the $f\left(
1370\right) $, $f\left( 1500\right) $, $f_{j}\left( 1710\right) $
by studying the mixing of these three states. The same situation
is also in the case of identification of the $2^{++}$ glueball.
Some progress has been made recently in the $0^{++}$ scalar and
$2^{++}$ tensor glueball sector, where both experimental and QCD
lattice simulation results seem to converge.

In this paper we will study the problem of glueball mass spectra
using the relativistic Dirac equation, which was used to the
calculations of mesons spectra, and the Breit-Fermi approach which
was also used to obtaining the hyperfine splitting in different
mesons. Our approach is based on the assumption that the nature
and the forces between two gluons are the same that between two
quarks, and because that the chosen approaches in both cases must
be similar. In our previous papers we obtained a good description
of meson spectra, fine and hyperfine splitting in relativistic
Dirac approach with screened potential. Because that we were able
to calculate the glueball masses and decays in the same approach,
with the same parameters.

\section{Method Calculation of Glueballs states}

For calculation of glueball mass-spectrum we used Dirac equation
with mixing Lorentz structure of potential for glueballs with spin
averaged masses

\begin{equation}
\begin{array}{c}
-\frac{dF_{ij}\left( r\right) }{dr}+\frac{k}{r}F_{ij}\left( r\right) =\left[
E-m-V\right] G_{ij}\left( r\right) \\
\frac{dG_{ij}\left( r\right) }{dr}+\frac{k}{r}G_{ij}\left( r\right) =\left[
E+m-V\right] F_{ij}\left( r\right)
\end{array}
\label{f1}
\end{equation}

where

\begin{equation}
\overrightarrow{\sigma }\overrightarrow{\ell }=\left(
\begin{array}{l}
j-\frac{1}{2} \\
-j-\frac{3}{2}
\end{array}
\right) =-\left( 1+k\right).  \label{f2}
\end{equation}

Let us write to equation all possible potential type

\[
i\frac{\partial \Psi \left( r,t\right) }{\partial t}=
\]

\begin{equation}
=\left( \frac{\alpha }{i}\nabla +\beta \left[ m+V^{S}\left(
r\right) +\gamma ^{5}V^{PS}\left( r\right) +\gamma ^{\mu }V_{\mu
}^{V}\left( r\right) +\gamma ^{\mu }\gamma ^{5}V_{\mu }^{A}\left(
r\right) +\sigma ^{\mu \eta }V_{\mu \eta }^{T}\right] \right).
\label{f3}
\end{equation}

The vector and scalar part we present in the following form:

\begin{equation}
\begin{array}{l}
V_{V}\left( r\right) =V_{OGE}\left( r\right) +\varepsilon V_{conf.}\left(
r\right) \\
V_{S}\left( r\right) =\left( 1-\varepsilon \right) V_{conf.}\left( r\right)
\end{array}
\label{f4}
\end{equation}

and substituted it to $\left( \ref{f1}\right)$

\begin{equation}
\begin{array}{l}
\left( E-V_{OGE}\left( r\right) -V_{conf.}\left( r\right) -m\right) F\left(
r\right) =-\frac{k}{r}G\left( r\right) -\frac{dG\left( r\right) }{dr} \\
\left( E-V_{OGE}\left( r\right) +\left( 1-2\varepsilon \right)
V_{conf.}\left( r\right) +m\right) G\left( r\right) =-\frac{k}{r}F\left(
r\right) +\frac{dF\left( r\right) }{dr}
\end{array}
\label{f5}
\end{equation}

\[
\left( E-V_{OGE}\left( r\right) -V_{conf.}\left( r\right) -m\right) F\left(
r\right) =
\]

\begin{equation}
=-\frac{k}{r}\frac{-\frac{k}{r}F\left( r\right) +\frac{dF\left(
r\right) }{dr}}{\left( E-V_{OGE}\left( r\right) +\left(
1-2\varepsilon \right) V_{conf.}\left( r\right) +m\right)},
\label{f6}
\end{equation}

\[
\left( E-V_{OGE}\left( r\right) -V_{conf.}\left( r\right) -m\right) F\left(
r\right) =
\]

\begin{equation}
=-\frac{k}{r}\frac{-\frac{k}{r}F\left( r\right) +\frac{dF\left(
r\right) }{dr}}{\left( E-V_{OGE}\left( r\right) +\left(
1-2\varepsilon \right) V_{conf.}\left( r\right) +m\right)}.
\label{f7}
\end{equation}

We obtain for the small component of the wave function $G\left(
r\right)$:

\[
\frac{dG\left( r\right) }{dr}=-\left( E-V_{OGE}\left( r\right)
-V_{conf.}\left( r\right) -m\right) F\left( r\right) -
\]

\begin{equation}
-\frac{k}{r}\left[ \frac{-\frac{k}{r}F\left( r\right)
+\frac{dF\left( r\right) }{dr}}{E-V_{OGE}\left( r\right) +\left(
1-2\varepsilon \right) V_{conf.}\left( r\right) +m}\right],
\label{f8}
\end{equation}

\[
\left[ \left( 1-2\varepsilon \right) V_{conf.}\left( r\right) -V_{OGE}\left(
r\right) \right] G\left( r\right) +
\]

\begin{equation}
+\left( E+\left( 1-2\varepsilon \right) V_{conf.}\left( r\right)
-V_{OGE}\left( r\right) +m\right) \frac{dG\left( r\right) }{dr}=  \label{f9}
\end{equation}

\[
=\frac{k}{r^{2}}F\left( r\right) -\frac{k}{r}F^{\prime }\left(
r\right) +F^{\prime \prime }\left( r\right).
\]
Into the formula $\left( \ref{f5}\right)$ we substitute $G\left(
r\right)$, so we obtain the second order equation for the wave
function $F\left( r\right)$:

\[
F^{\prime \prime }\left( r\right) -\frac{\left[ \left(
1-2\varepsilon \right) V_{conf.}\left( r\right) -V_{OGE}\left(
r\right) \right] ^{\prime }}{ \left( E+\left( 1-2\varepsilon
\right) V_{conf.}\left( r\right) -V_{OGE}\left( r\right) +m\right)
}F^{\prime }\left( r\right) +
\]

\begin{equation}
+ \left\{ \frac{k}{r^{2}}-\frac{k^{2}}{r^{2}}+\frac{\left[ \left(
1-2\varepsilon \right) V_{conf.}\left( r\right) -V_{OGE}\left(
r\right) \right] ^{\prime }}{\left( E+\left( 1-2\varepsilon
\right) V_{conf.}\left( r\right) -V_{OGE}\left( r\right) +m\right)
}\frac{k}{r}+\right.  \label{f10}
\end{equation}

\[
+\left( E+\left( 1-2\varepsilon \right) V_{conf.}\left( r\right)
-V_{OGE}\left( r\right) +m\right) \left( E-V_{OGE}\left( r\right)
-V_{conf.}\left( r\right) -m\right) \left. {}\right\} F\left(
r\right)=0.
\]

As for the interaction potential we choose the same screened
potential as in the case of Breit-Fermi equation. With the same
interaction potential we obtained good results in meson
spectroscopy, and because that we try to apply this approach to
glueballs. In the frame of Breit-Fermi approach the spin - spin
interaction term is

\begin{equation}
V_{SS}=\frac{2}{3m_{q_{1}}m_{q_{2}}}\overrightarrow{S_{1}}\overrightarrow{S_{2}}\Delta
V_{V}.  \label{f11}
\end{equation}

We suggest for calculation to use configuration interaction
approach, which was very successfully applied in atomic physics.

\section{Results}

In the Table 2 we show the results obtained in potential model for
spin average glueball masses in comparison with the data of other
authors. As it is seen from this table for the lowest glueball
state our result is close for to the results of other authors. In
our model the first and second orbital as well radial excited
states are lower then in the lattice calculation \cite{1},
\cite{2}. The same situation is also in the Table 3, where we show
the results which incorporated the spin-spin effects. These
probably mean, that the strong interaction between two gluons in
glueball more strong than between two quarks in mesons. It is
interesting to note, that to a same conclusion come J. Cui and H.
Jin \cite{3} during calculation of glueball masses, using
Bethe-Salpeter equation.

\medskip

Table 1. Spin averaged glueball masses from Dirac equation with
$\left( V_{0}=0.586GeV\right)$.

\medskip

\begin{tabular}{|ccccc|}
\hline \multicolumn{1}{|c|}{
\begin{tabular}{c}
$Quantum$ \\
$numbers$%
\end{tabular}
} & \multicolumn{1}{c|}{
\begin{tabular}{c}
$Our$ $results$ \\ $M\left( GeV\right)$
\end{tabular}
} & \multicolumn{1}{c|}{
\begin{tabular}{c}
$Nonperturb.method$ \\ $\left[ 4\right]$
\end{tabular}
} & \multicolumn{1}{c|}{
\begin{tabular}{c}
$Lattice$ $data$ \\ $\left[ 1\right]$
\end{tabular}
} &
\begin{tabular}{c}
$Lattice$ $data$ \\ $\left[ 2\right]$
\end{tabular}
\\ \hline
\multicolumn{1}{|c|}{$\ell =0,n_r=0$} &
\multicolumn{1}{c|}{$4.68$} & \multicolumn{1}{c|}{$4.68$} &
\multicolumn{1}{c|}{$4.66$} & $4.55$ \\ \hline
\multicolumn{1}{|c|}{$\ell =1,n_r=0$} &
\multicolumn{1}{c|}{$5.84$} & \multicolumn{1}{c|}{$6.0$} &
\multicolumn{1}{c|}{$6.36$} & $6.1$ \\ \hline
\multicolumn{1}{|c|}{$\ell =0,n_r=1$} &
\multicolumn{1}{c|}{$6.22$} & \multicolumn{1}{c|}{$6.0$} &
\multicolumn{1}{c|}{$6.68$} & $6.45$ \\ \hline
\multicolumn{1}{|c|}{$\ell =2,n_r=0$} &
\multicolumn{1}{c|}{$6.69$} & \multicolumn{1}{c|}{$7.0$} &
\multicolumn{1}{c|}{$9.0$} & $7.7$ \\ \hline
\multicolumn{1}{|c|}{$\ell =1,n_r=1$} &
\multicolumn{1}{c|}{$6.96$} & \multicolumn{1}{c|}{$8.0$} &
\multicolumn{1}{c|}{} &  \\ \hline
\end{tabular}

\bigskip

Table 2. Glueball masses from Breit-Fermi approach in comparison
with other authors.

\medskip

\begin{tabular}{|cccccccc|}
\hline \multicolumn{1}{|c|}{$J^{PC}$} & \multicolumn{1}{c|}{$our$
$results,M\left( GeV\right) $} & \multicolumn{1}{c|}{$ref$.$\left[
4\right] $} & \multicolumn{1}{c|}{$ref$.$\left[ 2\right] $} &
\multicolumn{1}{c|}{$ref$.$ \left[ 1\right] $} &
\multicolumn{1}{c|}{$ref$.$\left[ 5\right] $} &
\multicolumn{1}{c|}{$ref$.$\left[ 6\right] $} & $ref$.$\left[
7\right] $ \\ \hline \multicolumn{1}{|c|}{$0^{++}$} &
\multicolumn{1}{c|}{$1.70$} & \multicolumn{1}{c|}{$1.58$} &
\multicolumn{1}{c|}{$1.73$} & \multicolumn{1}{c|}{$1.74$} &
\multicolumn{1}{c|}{$1.645$} & \multicolumn{1}{c|}{$1.686$} &
$1.659$ \\ \hline \multicolumn{1}{|c|}{$2^{++}$} &
\multicolumn{1}{c|}{$2.19$} & \multicolumn{1}{c|}{$2.59$} &
\multicolumn{1}{c|}{$2.40$} & \multicolumn{1}{c|}{$2.47$} &
\multicolumn{1}{c|}{$2.337$} & \multicolumn{1}{c|}{$2.380$} &
$2.304$ \\ \hline \multicolumn{1}{|c|}{$3^{++}$} &
\multicolumn{1}{c|}{$3.24$} & \multicolumn{1}{c|}{$3.58$} &
\multicolumn{1}{c|}{$3.69$} & \multicolumn{1}{c|}{$4.3$} &
\multicolumn{1}{c|}{} & \multicolumn{1}{c|}{} &
\\ \hline
\end{tabular}

\bigskip

The Decays of Glueballs to two light mesons:

$\Gamma \left( 1^{3}S_{1}\rightarrow XX\right)=55MeV$,

$\Gamma \left( 2^{3}S_{1}\rightarrow XX\right)=24MeV$,

$\Gamma \left( 3^{3}S_{1}\rightarrow XX\right)=15MeV$.

\section{Discussion and conclusion}

As it is note by Minkowski and Ochs \cite{8}, \cite{11} in the
gluon jet in case of triplet neutralization the leading particles
are the hadrons formed by the primary gluon and soft qq pairs. If
there are hybrid mesons they may be formed in the fast color
singlet qqg system alternatively the fast hadrons are ordinary qq
mesons formed at the end of the parton cascade after having
absorbed all gluon energy. If the octet mechanism is at work the
leading gluon may also form a glueball. On the other hand in the
quark jet neither the hybrid nor the glueball will be leading if
the leading quark is only neutralized in color by a soft
antiquark.

\medskip

Table 3. Production of leading hadrons in the jet according
\cite{8}.

\medskip

\begin{tabular}{|l|l|l|l|l|}
\hline
& $neutralization$ & $qq$ & $hybrid$ & $glueball$ \\ \hline
$quark$ $jet$ & $triplet$ & $yes$ & $no$ & $no$ \\ \hline
$gluon$ $jet$ & $triplet$ & $yes$ & $yes$ & $no$ \\ \hline
& $octet$ & $no$ & $no$ & $yes$ \\ \hline
\end{tabular}

\bigskip

As it is note by Burakovsky \cite{9} it is widely believed that
pseudoscalar mesons are the Goldstone bosons of broken
SU(3)$\times $SU(3) chiral symmetry of QCD, and that they should
be massless in the chirally symmetric phase. Because that it is
not clear how the resonance spectrum would be suitable for the
description of the pseudoscalar mesons.

It is firmly established that the lightest glueball state is the scalar
glueball. According to Burakovsky, who obtained the glueball masses from the
Regge phenomenology, with linear Regge trajectory with negative intercept

\[
M^{2}\left( 0^{++}\right) =\frac{3}{\sqrt{2}}M\left( \rho \right)
\]
and
\[
M^{2}\left( 2^{++}\right) =\sqrt{2}M\left( 0^{++}\right)
\]

So $M\left( 0^{++}\right)=1620$ $MeV$, $M\left(
2^{++}\right)=2290$ $MeV$.

Tensor glueball is the lowest resonance lying on the Pomeron
trajectory with unit intercept. As for this glueball state in PDG
corresponds three candidates in this mass region $f_{j}\left(
2220\right)$, $J=2$ or $4$, $f_{2}$ $\left( 2300\right)$ and
$f_{2}\left( 2340\right)$.

As to the question of hybrids. If hybrid spectroscopy works like
conventional mesons spectroscopy or glueball spectroscopy than as
shown by Toussaint \cite{10}, we expect that the lowest multiplet
of hybrids contains $1^{-+}$, $1^{-}$, $0^{-+}$, $2^{-+}$
particles both isovector and isoscalar which will be split by
color hyperfine interaction. In this paper we restrict ourselves
only with the calculations of glueball masses.

Using a non-perturbative method based on asymptotic behavior of
Wilson loops A. Kaidalov and Yu. Simonov calculated the masses of
glueballs and corresponding Regge trajectories. These results are
shown in the Tables 1, 2.

The difficulties in identification of the glueball states, and the
poor experimental data makes the investigation of the problem for
glueballs actually. We try to obtain the mass spectrum of
glueballs and the decay width in potential model approach. Using
the same method and the same parameters as in the case of meson
spectra calculation we obtained for glueball masses a good
agreement with the possible experimental data. We obtained the
results for spin average glueball states and the results which
take into account the spin-spin interaction. Moreover, our
results, as well as other authors results, shown, that the lowest
scalar glueball mass is $M\left( 0^{++}\right)=1700$ $MeV$, and
the tensor glueball amass is $M\left( 2^{++}\right)=2290$ $MeV$.
Other results in Tables 1, 2 are predictions.

\end{document}